\documentclass[11pt,a4paper]{article}
\usepackage{jheppub}
\usepackage{graphicx}
\usepackage{amssymb,amsmath,amsbsy}

\author[a]{Sacha Davidson,}
\author[b,c]{Ricardo Gonz\'{a}lez Felipe,}
\author[d]{H.~Ser\^{o}dio,}
\author[b,c]{Jo\~{a}o P.~Silva}
\affiliation[a]{IPNL, Universit\'{e} de Lyon, Universit\'{e} Lyon 1, CNRS/IN2P3, 4
rue E. Fermi 69622 Villeurbanne cedex, France}
\affiliation[b]{Instituto Superior de Engenharia de Lisboa - ISEL, 1959-007
Lisboa, Portugal}
\affiliation[c]{Centro de F\'{\i}sica Te\'{o}rica de Part\'{\i}culas (CFTP), Instituto
Superior T\'{e}cnico, Universidade T\'{e}cnica de Lisboa, 1049-001 Lisboa, Portugal}
\affiliation[d]{Departament de F\'{\i}sica Te\`{o}rica and IFIC, Universitat de
Val\`{e}ncia-CSIC, E-46100, Burjassot, Spain}
\emailAdd{s.davidson@ipnl.in2p3.fr}
\emailAdd{ricardo.felipe@ist.utl.pt}
\emailAdd{hugo.serodio@ific.uv.es}
\emailAdd{jpsilva@cftp.ist.utl.pt}

\title{Baryogenesis through split Higgsogenesis}

\abstract{We study the cosmological evolution of asymmetries in the two-Higgs
doublet extension of the Standard Model, prior to the electroweak phase
transition. If Higgs flavour-exchanging interactions are sufficiently slow,
then a relative asymmetry among the Higgs doublets corresponds to an
effectively conserved quantum number. Since the magnitude of the Higgs
couplings depends on the choice of basis in the Higgs doublet space, we
attempt to formulate basis-independent out-of-equilibrium conditions. We show
that an initial asymmetry between the Higgs scalars, which could be generated
by CP violation in the Higgs sector, will be transformed into a baryon
asymmetry by the sphalerons, without the need of $B-L$ violation. This novel
mechanism of baryogenesis through (split) Higgsogenesis is exemplified with
simple scenarios based on the out-of-equilibrium decay of heavy singlet
scalar fields into the Higgs doublets.}

 \arxivnumber{}

\begin{document}
\maketitle

\section{\label{sec:intro}Introduction}

There is at least one Higgs boson~\cite{LHC}; maybe there are more.
Multi-Higgs doublet models contain new sources of CP violation, which is one
of the required ingredients~\cite{Sakharov} for baryogenesis. It is therefore
interesting to consider whether CP violation from the Higgs sector could be
used to generate the baryon asymmetry of the Universe~\cite{BAU,BAU2}. This
can occur in electroweak baryogenesis scenarios \cite{BAU@EPT}; here we are
interested in asymmetries produced before the electroweak phase transition
(EWPT).

In this paper we consider two-Higgs doublet models (2HDM)~\cite{hhg,2HDM}. If
interactions which exchange Higgs flavour are sufficiently weak, then the two
populations of Higgs fields could  contain independent asymmetries in the
early Universe. Since at least one of the Higgs must couple to Standard Model
(SM) fermions, its asymmetry is redistributed among other SM particles by
Yukawa interactions,  prior to the electroweak phase transition.
However, hypercharge neutrality of the Universe relates the asymmetries among
all charged particles. This implies that a relative asymmetry among the Higgs
scalars, generated by out-of-equilibrium CP-violating processes in the Higgs
sector, could be transformed into a baryon asymmetry in the presence of
($B+L$)-violating sphalerons~\cite{sphalerons}. The interest of such
baryogenesis scenarios is that they require no $B$ or $L$-violating
interactions beyond the non-perturbative sphalerons of the SM, relying only on
CP violation in an extended Higgs sector.

The issue of ``basis-independence'' is of particular
importance~\cite{Lavoura:1994fv,DH}. The point is that physical observables
cannot depend on a basis choice in the Lagrangian ---one may ask, for
instance, what $\phi_1$ and $\phi_2$ are in the 2HDM. Clearly, the survival of
a relative asymmetry between the $\phi_1$'s and $\phi_2$'s in early Universe
will depend on the speed of interactions that exchange $\phi_1$ with $\phi_2$.
But the pertinent coupling constants naively appear to depend on the choice of
$\phi_1$ and $\phi_2$. We show that such washout interactions are controlled
by the misalignment among different couplings, and can be parameterised in a
basis-independent way.

The paper is organized as follows. A compendium of relevant results for the
2HDM is given in section~\ref{sec:2HDM}, followed by some estimates for
interaction rates in the early Universe. Section~\ref{sec:TE} constrains
parameters of the Higgs potential by requiring Higgs flavour exchange to be
out of equilibrium. In the second part of this section, we discuss the
basis-independence of these bounds. In Section~\ref{sec:chemeq} we derive the
equations of chemical equilibrium~\cite{KS},  which relate the asymmetries
among SM particles and Higgs fields, due to the interactions which are in
equilibrium. As a result, a nonvanishing equilibrium baryon asymmetry is
obtained in the presence of a relative Higgs asymmetry, even with $B-L$
conservation. Simple scenarios based on the out-of-equilibrium decay of
singlet scalar fields into Higgs doublets are presented in
Section~\ref{sec:scenarios}. Finally, our conclusions are summarized in
Section~\ref{sec:summary}.

\section{The 2HDM at finite temperature}
\label{sec:notn}

\subsection{Notation and review}
\label{sec:2HDM}

The interaction Lagrangian for  the  general 2HDM~\cite{hhg,2HDM} consists of
a scalar potential plus Yukawa coupling terms. The most general gauge
invariant scalar potential can be written as
\begin{eqnarray}
\label{pot}
V&=& m_{11}^2\phi_1^\dagger\phi_1+m_{22}^2\phi_2^\dagger\phi_2
-[m_{12}^2\phi_1^\dagger\phi_2+{\rm h.c.}]\nonumber\\[6pt]
&&\quad +\frac{1}{2}\lambda_1(\phi_1^\dagger\phi_1)^2
+\frac{1}{2}\lambda_2(\phi_2^\dagger\phi_2)^2
+\lambda_3(\phi_1^\dagger\phi_1)(\phi_2^\dagger\phi_2)
+\lambda_4(\phi_1^\dagger\phi_2)(\phi_2^\dagger\phi_1)
\nonumber\\[6pt]
&&\quad +\left[\frac{1}{2}\lambda_5(\phi_1^\dagger\phi_2)^2
+\lambda_6(\phi_1^\dagger\phi_1)(\phi_1^\dagger\phi_2)
+\lambda_7(\phi_2^\dagger\phi_2)(\phi_1^\dagger\phi_2)+{\rm H.c.}\right]\,,
\end{eqnarray}
where $\phi_1$ and $\phi_2$ are two complex SU(2)$_{L}$ doublet scalar fields
of unit hypercharge; $m_{11}^2$, $m_{22}^2$, and $\lambda_1\ldots\lambda_4$
are real parameters, while $m_{12}^2$ and $\lambda_5,\lambda_6,\lambda_7$ can
be complex. In  general, both $\phi_1$ and $\phi_2$ can have Yukawa couplings
to all the SM fermions. The Yukawa interactions are
\begin{eqnarray}
- {\cal L}_Y &=&
\overline{Q_L}\, (\mathbf{\Gamma}_1 \phi_1 + \mathbf{\Gamma}_2 \phi_2)\, d_R
+ \overline{Q_L}\, (\mathbf{\Delta}_1 \tilde{\phi}_1 + \mathbf{\Delta}_2
\tilde{\phi}_2)\, u_R +
\nonumber\\
&& +\, \overline{L_L}\, (\mathbf{\Pi}_1 \tilde{\phi}_1 + \mathbf{\Pi}_2
\tilde{\phi}_2)\, \ell_R + \textrm{H.c.}, \label{Yuk_Z2}
\end{eqnarray}
where $Q_L = (u_L, d_L)^T$ ($u_R$ and $d_R$) is a vector in the 3-dimensional
generation space of left-handed quark doublets (right-handed charge $+2/3$ and
$-1/3$ quarks). Accordingly, $\mathbf{\Gamma}_1$, $\mathbf{\Gamma}_2$,
$\mathbf{\Delta}_1$, and $\mathbf{\Delta}_2$ are $3 \times 3$ matrices in the
respective quark generation spaces. Similarly, $L_L = (\nu_L, \ell_L)^T$ and
$\ell_R$ are vectors in the 3-dimensional generation space of left-handed
lepton doublets and right-handed charged leptons, respectively, while
$\mathbf{\Pi}_1$ and $\mathbf{\Pi}_2$ are $3 \times 3$ matrices. For
simplicity, we assume that there are no right-handed neutrino fields.

Under global SU(2) transformations in $(\phi_1, \phi_2)$ space, the kinetic
terms of the Higgs doublets are invariant, whereas the parameters of the
scalar potential (and the Yukawa couplings) will be modified.  Such basis
transformations in the Lagrangian cannot affect observables,  so the numerical
value of the parameters in Eq.~(\ref{pot}) is only meaningful when the basis
is specified. Three obvious Higgs basis choices can be envisaged:
\begin{itemize}
\item   $m_{12}^2 = 0$ basis, where we put a tilde on the parameters
    ($\tilde{\lambda}_i, \tilde{m}_{ii}^2$, $\tilde{y}^f_i$),
\item symmetry basis, where the parameters are lower case with a prime
    ($\lambda'_i, m_{ij}^{'2}$, ${y'}^f_i$),
\item (thermal) mass eigenstate basis, where the parameters are uppercase
    ($\Lambda_i, M_{ij}$, ${Y}^f_i$).
\end{itemize}
Here, $\tilde{y}^f_i$, ${y'}^f_i$ and ${Y}^f_i$ denote the Yukawa matrices of
the SM fermions $f$ interacting with the Higgs $i$, in the corresponding Higgs
basis (so $y^u_i = (\Delta_1,\Delta_2)$, and so on).

Since our goal is to store an asymmetry between the Higgs populations prior to
the EWPT, interactions which exchange $\phi_1 \leftrightarrow \phi_2$ must be
small (see next section). We refer to such interactions as (Higgs)
flavour-exchanging processes. For instance, in the $m_{12}^2 = 0$ basis, the
offending parameters from the potential are $\tilde{\lambda}_5$,
$\tilde{\lambda}_6$ and $\tilde{\lambda}_7$. In the Yukawa sector,
interactions  of both Higgs doublets to either quarks or leptons will be
strongly constrained. This is because a relative asymmetry in the two Higgs
populations should be preserved, so the two Higgs fields cannot both share
their asymmetry with the same fermions.

Some of the undesirable couplings can be suppressed by imposing a  discrete
$Z_2$  symmetry
\begin{equation} \phi_1 \to -\phi_1\,,\quad \phi_2 \to \phi_2\,.\label{Z2}
\end{equation}
In the basis where the symmetry has the above form, it implies $m'_{12}
=\lambda'_6 =   \lambda'_7 = 0$, so the Higgs sector contains no explicit CP
violation, because the phase of $\lambda'_5$ can be rotated away by a phase
choice of the Higgs fields. If both scalar fields couple to fermions of the
same charge, then there will be flavour changing neutral scalar interactions,
which are strongly constrained by experiment. The undesirable Yukawa
interactions can be removed by extending the $Z_2$ symmetry of Eq.~(\ref{Z2})
to the fermion sector, so that each fermion charge sector only couples to one
of the Higgs scalars. The four ways to implement this symmetry are shown in
Table~\ref{table1}.
%
\begin{table*}[t!]
\centering

\begin{tabular}{|c|cc|}
\hline
Model type & $\phi_1$ & $\phi_2$\\
\hline
Type I &   & $u$, $d$, $\ell$\\
Type II & $d$, $\ell$  & $u$ \\
Type X  &  $\ell$ & $u$, $d$ \\
Type Y &  $d$  & $u$, $\ell$\\
\hline
\end{tabular}
\caption{The four types of $Z_2$ models and the corresponding Higgs couplings
to fermions. Type X is also known as ``lepton specific'', and type Y as
``flipped''. In the usual $\phi_{u,d}$ notation, $\phi_u = \phi_2$
always.}
\label{table1}
\end{table*}

The discovery of a 125 GeV scalar at LHC places constraints on the 2HDM
parameter space, studied so far in the context of a $Z_2$ symmetry
($\lambda'_6 = \lambda'_7 = 0$), occasionally exact ($m'_{12} =
0$)~\cite{reviews}. After electroweak symmetry breaking, the neutral
components of the scalar fields acquire the vacuum expectation values (VEVs)
$\langle \phi_1^0 \rangle = v_1$ and $\langle \phi_2^0 \rangle = v_2$, where
$v = (v_1^2 + v_2^2)^{1/2} \simeq 174$~GeV. Of the eight components in the
two Higgs doublets, three provide longitudinal components to $W^\pm$ and $Z$,
two create a $H^\pm$ pair, two yield the Higgs scalar ($m_h = 125$ GeV) and
another neutral scalar ($H$), and the last gives a pseudoscalar ($A$). These
masses and the VEVs are related to the parameters of the potential. In
particular, \textit{if both $v_1 \neq 0$ and $v_2 \neq 0$}, then the
stationarity conditions can be used to write the pseudoscalar mass
\begin{equation}
m_A^2 = \frac{v^2}{v_1 v_2}\, m^{'2}_{12} - 2\lambda'_5\, v^2.
\end{equation}
This shows that requiring small $\phi_1 \leftrightarrow \phi_2$ exchanges
through $m'_{12} \sim 0$ and $\lambda'_5 \sim 0$, leads to $m_A \sim 0$,
unless $v_1 v_2 \sim 0$. This occurs because, in the $m'_{12} = \lambda'_5 =
\lambda'_6 = \lambda'_7 = 0$ limit, the potential in Eq.~\eqref{pot} has a
global $U(1)$ symmetry, which is broken by $v_1 v_2 \neq 0$, with the
consequent appearance of a massless Goldstone boson ($m_A=0$).

One solution is to consider the inert model~\cite{inert}, which is a Type I
2HDM with exact $Z_2$ and $v_1 = 0$\footnote{In the usual notation for the
inert doublet model, only $\phi_1$ couples to fermions, while $v_2=0$. In the
notation used here, the role of $\phi_1$ and $\phi_2$ are reversed, implying
the changes $m'_{11} \leftrightarrow m'_{22}$ and $\lambda'_1 \leftrightarrow
\lambda'_2$.}. In that case,
\begin{equation}
m_A^2 = m^{'2}_{22} + (\lambda'_3 + \lambda'_4- \lambda'_5)\, v^2.
\label{mA_inert}
\end{equation}
This mass can be kept nonzero, even if $\lambda'_5 = 0$, because the vacuum
with $v_1=0$ does not break the global $U(1)$. The only consequence of
$\lambda'_5 = 0$ is $m_H=m_A$. Because the inert $\phi_1$ does not couple to
fermions, the lightest particle is a candidate for dark matter. A series of
very clear analyses of this model, including constraints from both LHC and
WMAP, have been performed by the Warsaw group~\cite{warsaw}. They find large
regions of parameter space consistent with all known data, especially if the
$h \rightarrow \gamma \gamma$ signal is consistent with the SM
($R_{\gamma\gamma} \sim 1$). This is within the 2$\sigma$ ranges of current
ATLAS~\cite{Aad:2013wqa} and CMS~\cite{CMS_gg} data
\begin{eqnarray}
\textrm{ATLAS:}
&\ \ \ &
R_{\gamma\gamma} = 1.55^{+0.33}_{-0.28}\,,
\nonumber\\
\textrm{CMS:}
&\ \ \ &
R_{\gamma\gamma} = {0.78}^{+0.28}_{-0.26}\,.
\end{eqnarray}
Values of $R_{\gamma\gamma}$ larger than one restrict considerably the
parameter space.

An additional constraint imposed on the Higgs spectrum by our baryogenesis
scenario is that $\phi_1$ and $\phi_2$ should be present in the thermal bath
until the EWPT. If one of the $\phi_i$ is sufficiently heavy that its
population decays away prior to the EWPT, then the relative Higgs asymmetry
is lost.

\subsection{Thermal masses and interaction rates}

In this section we review interaction rates in a thermal bath.  The relevant
eigenbasis for the external leg particles should be the thermal mass
eigenstate basis, so we start by estimating the thermal mass matrix of the
Higgs scalars, at temperatures $T \gg |m_{ij}|$. At finite temperature, the
lowest order contribution to the mass-squared matrix is~\cite{CE}
\begin{equation}
m^2_{ij}(T) \simeq
\frac{\partial^2 V_{eff}(T, \phi_k)}{\partial\phi_i \partial\phi^*_j},
\label{defnmT}
\end{equation}
where  $V_{eff}(T, \phi_k)= V + V_T$ is the effective potential,
with $V$ given in Eq.~(\ref{pot}).
In the high temperature limit ($T \gg |m_{ij}|$),
\begin{equation}
V_T = \frac{T^2}{12} {\rm Tr} \big[ \mathbf{m}^2 \big] =
\frac{T^2}{24}\sum_{i=1,2} g_i M^2(\phi_i)\,,
\label{VT}
\end{equation}
where  $g_i = 4$  for a complex doublet field, and the trace is calculated
over the  $T=0$ scalar mass-squared matrix in background field, {\it  i.e.}
allowing $\phi_1$ and $\phi_2$ to have non-zero values so that $M(\phi_i)$
are field-dependent masses.

Neglecting zero-temperature loop contributions and finite-temperature fermion
and gauge contributions, we find
\begin{equation}
V_T =\frac{T^2}{12}\left\{
 3\lambda_1|\phi_1|^2  +(2 \lambda_{3} + \lambda_{4})(|\phi_2|^2+|\phi_1|^2)
+  3\lambda_2|\phi_2|^2
-
[3(\lambda_6 + 
 \lambda_7)\phi^{\dagger}_1 \phi_2
+{\rm H.c.}]\right\},
\end{equation}
where the $m_{ij}^2$ terms are dropped because they give no contribution to
Eq.~(\ref{defnmT}). For an arbitrary basis in the Higgs doublet space~\cite{Ivanov},
this gives the thermal mass-squared matrix
\begin{eqnarray}
\mathbf{m}^2(T)& \simeq&
\frac{T^2}{12}\left(
\begin{array}{cc}
3\lambda_1 + 2 \lambda_{3}   + \lambda_{4}&
-3\lambda_{6}   - 3\lambda_{7}\\
 -3\lambda^*_{6}   - 3\lambda^*_{7}
 & 3\lambda_2 + 2 \lambda_{3}   + \lambda_{4}\\
\end{array}
\right)
+
\left(
\begin{array}{cc}
 m_{11}^2&m_{12}^2 \\
m_{12}^{*2} &m_{22}^2
\end{array}
\right).
\label{mtherm}
\end{eqnarray}
Diagonalising this matrix gives the thermal mass eigenstate basis. In the
presence of a $Z_2$ symmetry, the thermal masses are simply given by
\begin{eqnarray}
m_{11}^2(T) &=& \frac{T^2}{12} (3 \lambda_1 + 2 \lambda_3 + \lambda_4) + m_{11}^2,
\nonumber\\
m_{22}^2(T) &=& \frac{T^2}{12} (3 \lambda_2 + 2 \lambda_3 + \lambda_4) + m_{22}^2,
\end{eqnarray}
so that no term of the type $m_{12}^2(T)\, \phi_1^\dagger \phi_2$ is
generated. Therefore, in the latter case, the only link between $\phi_1$ and
$\phi_2$ in the Higgs potential comes from the $\lambda_5$ term, both at zero
and at finite temperature.

We now review the assumptions and approximations involved in our estimates for
the interaction rates. We take ``thermal equilibrium" to describe a particle
species distributed following a Maxwell-Boltzmann distribution.  At
temperatures $T \ll m_{\rm GUT} \simeq 10^{16}$~GeV, this will be the case for
particles with SM gauge interactions. We define an interaction to be in
``chemical equilibrium'' if it is fast enough to impose relations among the
asymmetries in the participating particles. This will be the case if its
timescale, $1/\Gamma$, is much shorter than the age of the Universe $\sim
1/H$, i.e. $\Gamma \gg H$, where
\begin{equation}
H = \sqrt{\frac{4\pi^3 g_\ast}{45}} \frac{T^2}{m_P} \simeq \frac{
17\, T^2}{m_P}, \label{hubble}
\end{equation}
is the Hubble expansion parameter, $g_\ast$ is the number of
relativistic degrees of freedom ($g_\ast=107.75$ in the 2HDM) and $m_P = 1.22
\times 10^{19}$~GeV is the Planck mass.

We estimate the interaction rate $\Gamma = \gamma/n$, where $n_i   \simeq
g_iT^3/\pi^2$ is the equilibrium density of an incident (massless) particle,
and $\gamma$ is the interaction density.  For a process $ij \to mn$,  where
all the participating particles are in thermal equilibrium, $\gamma$  is the
thermally averaged scattering rate,
\begin{eqnarray} \label{gammarate}
\gamma(ij &\to& mn)
= \langle n_i n_j \sigma(i+j \to..) \rangle \nonumber\\
 && = \int d\Pi_i  d\Pi_j f^{eq}_i f^{eq}_j
\int |{\cal M}(i+j \rightarrow m+n )|^2
~(2\pi)^4{\delta}^4( p_i+p_j  -  p_m - p_n)~
d\Pi_m d\Pi_n \nonumber \\
&&= \frac{g_1 g_2 |\Lambda|^2\, T^4}{ 32\pi^5},
\label{gamlam}
\end{eqnarray}
where $g_i$ is the number of degrees of freedom of the particle in the  bath
(2 for a doublet), $d\Pi = \dfrac{d^3p}{2E(2 \pi)^3}$ is the relativistic
phase space, and $f^{eq}$ is the Maxwell-Boltzmann equilibrium distribution.
The last equality in Eq.~\eqref{gammarate} is the result for $|{\cal M}(i+j
\rightarrow m+n )|^2 = |\Lambda|^2$.

\section{Keeping Higgs flavour-exchanging interactions out of equilibrium}
\label{sec:TE}

We suppose that particle-antiparticle asymmetries in $\phi_1$ and $\phi_2$
were generated at some earlier epoch of the Universe. In
section~\ref{sec:scenarios}, we shall illustrate this in a simple framework.
We focus on the relative asymmetry between the two Higgs doublets:
\begin{equation}
\mathcal{Y}_{\Delta \phi_1} - \mathcal{Y}_{\Delta \phi_2} \equiv
\frac{n_{\phi_1}- n_{\overline{\phi}_1}}{s}- \frac{n_{\phi_2}-
n_{\overline{\phi}_2}}{s}, \label{asym}
\end{equation}
where $s$ is the entropy density of the Universe. We use the notation
$\mathcal{Y}_{\Delta X}$ for the asymmetry $\mathcal{Y}_X -
\mathcal{Y}_{\overline{X}}$, where $\mathcal{Y}_X = n_X/s$ is the comoving number
density.
This asymmetry will be conserved as long as Higgs flavour-exchanging
interactions are out of equilibrium. In this section, we identify these
interactions,  estimate the constraints on the couplings, and express these
bounds in some useful bases.

In the thermal mass eigenstate basis, the flavour-exchanging Higgs
interactions that must be out of equilibrium are mediated by the quartic
couplings $\Lambda_5$, $\Lambda_6$, and $\Lambda_7$. Requiring $ \Gamma \ll H$
at $T \simeq 100$~GeV, and using Eqs.~(\ref{hubble}) and (\ref{gamlam}),
implies
\begin{equation}
|\Lambda_n|
\lesssim {\rm few} \times 10^{-7},\quad n =5,6,7,
\label{bdL}
\end{equation}
to keep the Higgs asymmetries separate for temperatures down to the EWPT. This
condition applies in the thermal mass eigenstate basis; we translate it below
to other bases.

Higgs flavour could also be exchanged via Yukawa couplings, if both Higgs
doublets interact with the same fermions.  For simplicity, we only consider
the third generation of fermions. The $t$, $b$ and $\tau$ have Yukawa
interactions to both Higgs fields, so their Yukawa couplings are vectors in
Higgs doublet space, which we represent capitalised in the thermal mass
eigenstate basis. For instance, the top Yukawa coupling is $(Y^t_1, Y^t_2)$,
with $m_t = Y^t_1\, v_{1}^{T} +  Y^t_2\,v_{2}^{T}$, where the $v_{i}^{T}$ are
the zero-temperature Higgs VEVs in the thermal basis. The survival of the
relative Higgs asymmetry requires that the Yukawa interactions, between a
fermion species $f = t,b,\tau$ and one of the Higgs doublets, be out of
equilibrium:
\begin{equation}
\min_i \gamma(f+g/\gamma \to f+\phi_i) \ll n_{f} H.
\end{equation}
For instance, in the case of the top quark, this gives
\begin{equation}
 \theta_t^2 \frac{\alpha_s}{16\pi^2} \frac{m_t^2}{v^2}   \ll 17 \frac{T}{m_P},
\label{topbd}
\end{equation}
where $\theta_t$ is the rotation angle between the thermal mass eigenstate
Higgs basis and the eigenvector of the top Yukawa coupling (cf.
Appendix~\ref{sec:appendix}):
\begin{equation} \label{thetatapprox}
\theta_t \simeq \frac{|Y^t_1\,Y^t_2|}{\sqrt{|Y^t_1|^2 + |Y^t_2|^2}}\,
\simeq
\frac{\left|m_{12}^2(T)\right|}{\left|m_{11}^2(T) - m_{22}^2(T)\right|}
\,,
\end{equation}
where the last expression is in the Yukawa eigenbasis.

At $T \simeq 100$~GeV, Eq.~\eqref{topbd} requires
\begin{equation}\label{thetatbound}
\theta_t\lesssim {\rm few}\times 10^{-7}.
\end{equation}
Furthermore, using $|\lambda_6|, |\lambda_7| \lesssim$ few $\times 10^{-7}$ to
satisfy Eq.~(\ref{bdL}), and assuming $m_{11}^2$ and $m_{22}^2$ of the order
of the lightest Higgs mass, it follows from Eqs.~\eqref{mtherm}
and~\eqref{thetatapprox} that
\begin{equation}\label{m12bound}
|m_{12}^2| \lesssim (100\,\textrm{MeV})^2,
\end{equation}
in the Yukawa eigenbasis.

We remark that, in obtaining the bound (\ref{topbd}), we approximate
$m_t=(|Y^t_1|^2 + |Y^t_2|^2)^{1/2}\, v $,  that is, we neglect the
misalignment between the top Yukawa coupling vector and the  zero-temperature
VEVs. This could underestimate the magnitude of the Yukawa coupling (as arises
for the $b$ and $\tau$ in the large $\tan \beta$ limit of the supersymmetric
SM). Therefore, the interaction rates we obtain will be lower bounds. Similar
bounds apply to other fermions $f$, with the replacement $m_t \to m_f$ (and
$\alpha_s \to \alpha_{\rm QED}/4 $ for leptons). This leads to the bounds
\begin{eqnarray}\label{thetabtau}
\theta_b &\lesssim& 10^{-5} \to  10^{-7}\quad \text{for the b quark}, \nonumber\\
\theta_\tau &\lesssim& 10^{-4} \to  10^{-6}\quad \text{for the $\tau$-charged lepton}.
\end{eqnarray}
The weaker limit corresponds to $|Y^f_1|^2 +|Y^f_2|^2=(m_f/v)^2$ and the
stronger one to $|Y^f_1|^2 +|Y^f_2|^2 \sim 1$.

\subsection{Basis-independent conditions}
\label{sec:basis}

In this section, the conditions given in Eqs.~(\ref{bdL}) and (\ref{topbd}),
which ensure the survival of a relative Higgs asymmetry, are expressed in a
way which is independent of the (Higgs) basis transformation
\begin{equation}
\Phi \rightarrow \mathbf{U}\, \Phi,
\label{HBT}
\end{equation}
where $\Phi = ( \phi_1,\ \phi_2)^T$ and $\mathbf{U}$ is a $2 \times 2$ unitary
matrix in Higgs flavour space. In Refs.~\cite{Lavoura:1994fv,DH},
basis-independent combinations of potential parameters were constructed by
contracting the parameters with the Higgs VEVs (a vector in Higgs doublet
space). We will construct similar invariants here, but replacing the Higgs VEV
with the top Yukawa coupling, which is more relevant for our scenario, and is
also a vector in Higgs space (in the one generation approximation). Indeed,
one can combine the top Yukawa couplings in
\begin{equation}
- {\cal L}_Y
=
\overline{t_L} \left( y_1^t,\ y_2^t\right)\,
\left(
\begin{array}{cc}
\phi_1\\
\phi_2
\end{array}
\right)\, t_R
+ \textrm{h.c.},
\end{equation}
into a vector
\begin{equation}
\hat{y}^t
=
\frac{1}{\sqrt{|y^t_1|^2 +|y^t_2|^2}}
\left(
\begin{array}{c}
y^t_1\\
y^t_2
\end{array}
\right),
\label{hat_y}
\end{equation}
transforming as $\hat{y}^t \rightarrow \mathbf{U}\, \hat{y}^t$,
and its orthogonal
\begin{equation}
\hat{\epsilon}^t
=
\frac{1}{\sqrt{|y^t_1|^2 +|y^t_2|^2}}
\left(
\begin{array}{c}
- y^{t\,\ast}_2\\
y^{t\,\ast}_1
\end{array}
\right),
\label{hat_eps}
\end{equation}
transforming as $\hat{\epsilon}^t \rightarrow
[\textrm{det}\,\mathbf{U}]^{-1}\, \mathbf{U}\, \hat{\epsilon}^t$. In the top
basis (see Appendix~\ref{sec:appendix}), these vectors become
\begin{equation}
\hat{y}^t
=
\left(
\begin{array}{c}
1\\
0
\end{array}
\right),
\ \ \ \
\hat{\epsilon}^t
=
\left(
\begin{array}{c}
0\\
1
\end{array}
\right).
\label{topbasis}
\end{equation}
From Eq.~(\ref{topbd}), it is clear that the direction in Higgs space of the
top Yukawa coupling $\hat{y}^t$ should approximately correspond to $\phi_1$
or $\phi_2$ of the thermal mass eigenstate basis. We then simply impose the
bounds of Eqs.~(\ref{bdL}) and (\ref{topbd}) in the basis of
Eq.~(\ref{topbasis}).

It is convenient to introduce some notation patterned on Ref.~\cite{DH}. The
quartic Higgs interactions can be represented via a four-index tensor which
appears in the Lagrangian as $\frac{1}{2} Z_{a \bar{b} c \bar{d}}
\Phi_{\bar{a}}^\dagger \Phi_b \Phi_{\bar{c}}^\dagger \Phi_d$ where
$\Phi^\dagger = (\phi_1^\dagger, \phi_2^\dagger)$, and $a,b,c,d=1,2$. The
barred (unbarred) notation keeps track of which indices transform as
$\mathbf{U}^\dagger$ ($\mathbf{U}$), under the basis
transformation~\eqref{HBT}. The elements of $Z_{a \bar{b} c \bar{d}}$ are
\begin{eqnarray}
&& Z_{1\bar{1}1\bar{1}}=\lambda_1\,,\qquad\qquad \,\,\phantom{Z_{2\bar{2}2\bar{2}}=}
Z_{2\bar{2}2\bar{2}}=\lambda_2\,,\nonumber\\
&& Z_{1\bar{1}2\bar{2}}=Z_{2\bar{2}1\bar{1}}=\lambda_3\,,\qquad\qquad
Z_{1\bar{2}2\bar{1}}=Z_{2\bar{1}1\bar{2}}=\lambda_4\,,\nonumber \\
&& Z_{1\bar{2}1\bar{2}}=\lambda_5\,,\qquad\qquad \,\,\phantom{Z_{2\bar{2}2\bar{2}}=}
Z_{2\bar{1}2\bar{1}}=\lambda_5^*\,,\\
&& Z_{1\bar{1}1\bar{2}}=Z_{1\bar{2}1\bar{1}}=\lambda_6\,,\qquad\qquad
Z_{1\bar{1}2\bar{1}}=Z_{2\bar{1}1\bar{1}}=\lambda_6^*\,,\nonumber \\
&& Z_{2\bar{2}1\bar{2}}=Z_{1\bar{2}2\bar{2}}=\lambda_7\,,\qquad\qquad
Z_{2\bar{2}2\bar{1}}=Z_{2\bar{1}2\bar{2}}=\lambda_7^*\,.\nonumber
\label{znum}
\end{eqnarray}
By analogy with the invariants $|Z_5|$, $|Z_6|$, and $|Z_7|$ presented in
Ref.~\cite{DH}, the following basis invariant quantities can be constructed:
\begin{eqnarray}
|S_5|
& \equiv &
| Z_{a \bar{b} c \bar{d}}
\, \hat{y}_{\bar{a}}^{t \ast}
\, \hat{\epsilon}_{b}^{t}
\, \hat{y}_{\bar{c}}^{t \ast}
\, \hat{\epsilon}_{d}^{t} |,
\label{szvv5}
\nonumber\\
|S_6|
& \equiv &
| Z_{a \bar{b} c \bar{d}}
\, \hat{y}_{\bar{a}}^{t \ast}
\, \hat{y}_{b}^{t}
\, \hat{y}_{\bar{c}}^{t \ast}
\, \hat{\epsilon}_{d}^{t} |,
\label{szvv6}\\
|S_7|
& \equiv &
| Z_{a \bar{b} c \bar{d}}
\, \hat{y}_{\bar{a}}^{t \ast}
\, \hat{\epsilon}_{b}^{t}
\, \hat{\epsilon}_{\bar{c}}^{t \ast}
\, \hat{\epsilon}_{d}^{t} |.\nonumber
\label{szvv7}
\end{eqnarray}
These correspond to $|\lambda_5|$, $|\lambda_6|$,  and $|\lambda_7|$ in the
basis of Eq. (\ref{topbasis}) and, consequently, $|S_n| \lesssim {\rm few}
\times 10^{-7}$ to satisfy Eq.~(\ref{bdL}).

As seen in Appendix~\ref{sec:appendix}, the rotation angle between the thermal
mass eigenstate basis and the top Yukawa eigenbasis of Eq.~(\ref{topbasis})
can be written in the top basis as
\begin{equation}
|\tan{(2 \theta_{t})}|
=
\frac{2 |m_{12}^{t 2}|}{|m_{11}^{t 2} -  m_{22}^{t 2}|}
=
\frac{2
\left|m_{a \bar{b}}^{t 2}\ \hat{y}_{\bar{a}}^{t \ast}\ \hat{\epsilon}_b^t
\right| }{ \left|
m_{a \bar{b}}^{t 2}\ \hat{y}_{\bar{a}}^{t \ast}\ \hat{y}_b^t -
m_{a \bar{b}}^{t 2}\ \hat{\epsilon}_{\bar{a}}^{t \ast}\ \hat{y}_b^t
\right|},
\end{equation}
where the last expression is manifestly basis invariant and, according to
Eq.~\eqref{thetatbound}, $\theta_t \lesssim  {\rm few} \times 10^{-7}$ to
satisfy Eq.~(\ref{topbd}).

Finally, the misalignment angles between the top, bottom and tau eigenbases
can be formulated in basis-independent notation as
\begin{eqnarray}
{\rm min} {\Big\{} \hat{y}^b\cdot \hat{y}^t,
 \hat{y}^b\cdot \hat{\epsilon}^t  {\Big\}}  &\simeq & \theta_b \lesssim 10^{-5}
\to 10^{-7},
\nonumber\\
{\rm min} {\Big\{} \hat{y}^\tau\cdot \hat{y}^t,
 \hat{y}^\tau\cdot \hat{\epsilon}^t  {\Big\}}  &\simeq & \theta_\tau
 \lesssim 10^{-4}\to 10^{-6},
\end{eqnarray}
where $\hat{y}^b$ and  $\hat{y}^\tau$ are defined analogously to $\hat{y}^t$
in Eq.~\eqref{hat_y} and the upper bounds on the right-hand sides follow from
Eq.~\eqref{thetabtau}.

In summary, in a 2HDM prior to the EWPT, a relative Higgs asymmetry can
survive provided the Yukawa interactions and the Higgs potential have a
certain form. The Higgs potential parameters should satisfy the constraints
$|\lambda_n| \lesssim$ few $\times 10^{-7}\ (n=5,6,7)$, and $|m_{12}^2|
\lesssim $ (100 MeV)$^2$. In the same basis,  SM singlet  fermions of a given
charge (up-type quarks, down-type quarks, charged leptons) should interact
with approximately only one Higgs field, that is, the model should be of type
I, II, X, or Y.

\section{Chemical equilibrium relations}
\label{sec:chemeq}

Let us now study the redistribution of asymmetries in conserved quantum
numbers due to interactions in equilibrium. We neglect lepton flavour
asymmetries, so that the (exactly and effectively) conserved quantum numbers
are the hypercharge, $B-L$, and the relative Higgs asymmetry. Assuming that
the asymmetries in all species are small, they are related to the chemical
potential $\mu$ as
\begin{equation}
n_i - n_{\overline{i}} =
\frac{g_i T^2}{6} \, \mu_i \times
\left\{
\begin{array}{cl}
2 & \textrm{for bosons}\\
1 & \textrm{for fermions}
\end{array}
\right.,
\end{equation}
where $g_i$ is the number of degrees of freedom of the particle.

We take the thermal bath to contain the SM fermions and gauge bosons, and two
Higgs doublets.  We consider temperatures just prior to the EWPT, when all the
Yukawa interactions are in equilibrium, but gauge symmetries are unbroken, so
that the gauge bosons have zero chemical potential. Our aim is to investigate
whether a Higgs asymmetry, as given in Eq.~\eqref{asym}, can be used to
generate a baryon asymmetry.

If an interaction is in chemical equilibrium, then the sum of the chemical
potentials of the participating particles should vanish. The SM Yukawa
interactions impose the relations
\begin{eqnarray}
- \mu_q + \mu_{\phi_d} + \mu_{d_R} &=& 0,
\label{yuk_d}
\\
- \mu_q - \mu_{\phi_u} + \mu_{u_R} &=& 0,
\label{yuk_u}
\\
- \mu_\ell + \mu_{\phi_e} + \mu_{e_R} &=& 0,
\label{yuk_e}
\end{eqnarray}
where $\phi_d$, $\phi_u$ and $\phi_e$ denote the scalar that couples to the
down-type quarks, up-type quarks and charged leptons, respectively. Since in
the usual notation $\phi_u = \phi_2$, the various models in Table~\ref{table1}
differ by whether $\phi_d$ and/or $\phi_e$ coincide with $\phi_2$.

The electroweak sphalerons  impose
\begin{equation}
3 \mu_q  + \mu_\ell = 0,
\label{Ewk_sphal}
\end{equation}
while the QCD sphalerons lead to the chemical equilibrium condition
\begin{equation}
2 \mu_q - \mu_{u_R}  - \mu_{d_R} = 0.
\label{QCD_sphal}
\end{equation}
Adding Eqs.~\eqref{yuk_d}, \eqref{yuk_u}, and \eqref{QCD_sphal}, we find
\begin{equation}
\mu_{\phi_d} - \mu_{\phi_u} = 0.
\label{crucial}
\end{equation}
In type II and type Y models, where $\phi_2$ couples  to up-type quarks, and
$\phi_1$  to down-type quarks, this forces $\mu_{\phi_1} - \mu_{\phi_2}$ to
vanish. Therefore, in 2HDM of type II and type Y, a relative Higgs asymmetry
would be washed out. In contrast, in type I and type X models, $\phi_d =
\phi_2 = \phi_u$, Eq.~\eqref{crucial} is trivially satisfied, and thus
one can have $\mu_{\phi_1} - \mu_{\phi_2} \neq 0$. Next we show that, provided
a Higgs asymmetry was created in early Universe, it can be used to generate a
baryon asymmetry at later times.

The baryon and lepton number comoving asymmetries are given by
\begin{eqnarray}
\mathcal{Y}_{\Delta B} &=&N_g (
2 \mu_q + \mu_{u_R} + \mu_{d_R})\,\frac{T^2}{3s}= 4N_g \mu_q \frac{T^2}{3s},
\nonumber\\
\mathcal{Y}_{\Delta L} &=& N_g (2 \mu_\ell + \mu_{e_R})\,\frac{T^2}{3s} =
- N_g (9 \mu_q + \mu_{\phi_e})\,\frac{T^2}{3s},
\end{eqnarray}
where $N_g = 3$ is the number of generations, and
Eqs.~\eqref{yuk_e}-\eqref{QCD_sphal} have been used to rewrite the right-hand
sides of these expressions. As a result,
\begin{equation}
\mathcal{Y}_{\Delta B} - \mathcal{Y}_{\Delta L} =
N_g \left( 13\, \mu_q + \mu_{\phi_e} \right)\,\frac{T^2}{3s}.
\label{BmL}
\end{equation}

Finally, hypercharge (or, equivalently, electric) neutrality of the plasma
gives
\begin{eqnarray}
N_g &&\left(
- \mu_{e_R} - \mu_\ell + \mu_q + 2 \mu_{u_R} - \mu_{d_R}
\right)
+ 2 ( \mu_{\phi_1} + \mu_{\phi_2} )
\nonumber\\
=&& N_g \left(
8 \mu_q + \mu_{\phi_e} + 3 \mu_{\phi_2}
\right)
+ 2 ( \mu_{\phi_1} + \mu_{\phi_2} )=0.
\label{Qem}
\end{eqnarray}
The equilibrium  baryon asymmetry can then be written as a function of $B-L$
and the Higgs asymmetry:
\begin{eqnarray}
\mathcal{Y}_{\Delta B}
&=&
\frac{8}{23} (\mathcal{Y}_{\Delta B} - \mathcal{Y}_{\Delta L})
+ \frac{3}{46}
\left( \mathcal{Y}_{\Delta \phi_1} - \mathcal{Y}_{\Delta \phi_2} \right)
\hspace{5mm}
\textrm{for type I},
\label{BI_TypeI}
\\
\mathcal{Y}_{\Delta B}
&=&
\frac{8}{23} (\mathcal{Y}_{\Delta B} - \mathcal{Y}_{\Delta L})
- \frac{33}{92}
\left( \mathcal{Y}_{\Delta \phi_1} - \mathcal{Y}_{\Delta \phi_2} \right)
\hspace{5mm}
\textrm{for type X}.
\label{BI_TypeX}
\end{eqnarray}

Eqs.~\eqref{BI_TypeI} and \eqref{BI_TypeX} give a baryon asymmetry in the
presence of a relative Higgs asymmetry, even if $\mathcal{Y}_{\Delta B}
-\mathcal{Y}_{\Delta L}=0$. Thus, in these cases, the baryon asymmetry is due
exclusively to the initial imbalance between the asymmetry in $\phi_1$ and
the asymmetry in $\phi_2$. We dub this scenario as split Higgsogenesis. As
far as we know, this is a novel mechanism for baryogenesis. This is
reminiscent of certain asymmetric dark matter (DM) models~\cite{ADMrev},
where an asymmetry is generated in a new dark sector (which contains the DM
candidate), and then is shared with the SM fermions. The role that the Higgs
can play in transferring asymmetries between the SM fermions and the dark
sector has been recently emphasized in Ref.~\cite{ST}. However, $\phi_1$ does
not seem to be a successful asymmetric DM candidate in the simple model
discussed here.\footnote{Symmetric dark matter and electroweak baryogenesis
have been recently discussed in the inert doublet model in
Refs.~\cite{BAUIDM}.}

Let us assume that, indeed, $\mathcal{Y}_{\Delta B} =\mathcal{Y}_{\Delta L}$. Using
Eqs.~\eqref{BmL}-\eqref{BI_TypeX}, we find
\begin{eqnarray}
\mathcal{Y}_{\Delta B}
&=&
\frac{3}{46}
\left( \mathcal{Y}_{\Delta \phi_1} - \mathcal{Y}_{\Delta \phi_2} \right)
=
-\frac{6}{13}\, \mathcal{Y}_{\Delta \phi_2} = \frac{6}{79}\,  \mathcal{Y}_{\Delta \phi_1}
\hspace{6mm}
\textrm{for type I},
\label{B_TypeI}
\\*[2mm]
\mathcal{Y}_{\Delta B}
&=&
- \frac{33}{92}
\left( \mathcal{Y}_{\Delta \phi_1} - \mathcal{Y}_{\Delta \phi_2} \right)
=
-\frac{6}{13}\,  \mathcal{Y}_{\Delta \phi_1} = \frac{66}{41}\,  \mathcal{Y}_{\Delta \phi_2}
\hspace{5mm}
\textrm{for type X}.
\label{B_TypeX}
\end{eqnarray}
The type I model is particularly interesting because $\phi_1$ does not couple
to any fermion and could act as dark matter. In that case,
$\mathcal{Y}_{\Delta \phi_1}/\mathcal{Y}_{\Delta B} = 79/6$, so the DM scalar
should be lighter than the proton\footnote{ Alternatively, one could assume a
primordial $B-L$ asymmetry, no relative Higgs asymmetry, and let the
$\lambda_{5}$ coupling (allowed by the $Z_2$ symmetry which ensures DM
stability) to equilibrate the asymmetry between the two Higgs fields. Then,
Eq.~\eqref{Qem} yields a Higgs asymmetry smaller than the baryon asymmetry: $
\mathcal{Y}_{\Delta \phi_1} = \mathcal{Y}_{\Delta \phi_2} = -
\mathcal{Y}_{\Delta B}/8$. This corresponds to a scalar DM mass $\sim 20$
GeV, which is ruled out by the width of the $Z$ boson. Furthermore,
$\lambda_5$ would mediate DM-anti-DM oscillations which would wash out the
asymmetry.} to obtain $\Omega_{\textrm{DM}} \sim 5\, \Omega_{B}$. We recall
that the mass density of baryons in the Universe today, as inferred from WMAP
in the context of $\Lambda$CDM cosmology, is~\cite{Komatsu:2010fb}
\begin{equation}\label{nBobs}
 \frac{m_p}{\rho_c}( n_B - n_{\overline B})=
 \Omega_{B} h^2 = 0.02255 \pm 0.0054,
\end{equation}
or equivalently,
\begin{equation}\label{YBobs}
  \mathcal{Y}_{\Delta B} = (8.79\pm0.44)\times 10^{-11},
\end{equation}
where $m_p$ is the proton mass, $h \equiv H_0/(100\, $km s$^{-1}\,$Mpc$^{-1})
= 0.742 \pm 0.036$ is the present Hubble parameter, and $\rho_c = 3H_0^2/(8
\pi G)$ is the critical density of a spatially flat Universe. On the other
hand, for the $\Lambda$CDM cosmology with three light neutrinos, the cold dark
matter relic abundance is $\Omega_{\textrm{DM}} h^2 = 0.1126\pm
0.0036$~\cite{Komatsu:2010fb}, so that the ratio of dark matter particles to
baryons is $\mathcal{Y}_{\textrm{DM}}/\mathcal{Y}_{\Delta B} \sim 5\,
m_p/m_{\textrm{DM}}$.

\section{Simple split-Higgsogenesis scenarios}
\label{sec:scenarios}

Our goal in this section is to provide a few simple scenarios for
baryogenesis through split-Higgsogenesis, where the cosmological baryon
asymmetry could in principle be generated via the out-of-equilibrium decay of
heavy singlet scalars into Higgs doublets.

\subsection{One extra singlet scalar}

We consider an inert 2HDM extended by one real scalar singlet. The two Higgs
doublets, $\phi_1$ and $\phi_2$, and the singlet scalar $S$ transform under
$Z_2$ as
\begin{equation}
\phi_1 \rightarrow - \phi_1,\quad
\phi_2 \rightarrow \phi_2,\quad
S \rightarrow -S.
\label{Z2_2d1s}
\end{equation}
The $Z_2$-invariant Higgs potential can be written as
\begin{equation}
V = V_\phi + V_S + V_{S \phi},
\label{VPPS}
\end{equation}
where
\begin{eqnarray}\label{V_vp_Z2}
V_\phi
&=&
m_{11}^2\, |\phi_1|^2 + m_{22}^2\, |\phi_2|^2
+\, \tfrac{1}{2}\, \lambda_1\, |\phi_1|^4 + \tfrac{1}{2}\, \lambda_2\,
|\phi_2|^4 + \lambda_3\, |\phi_1|^2 |\phi_2|^2 + \lambda_4\, (\phi_1^\dagger
\phi_2)(\phi_2^\dagger \phi_1)
\nonumber\\
&&
+\, \tfrac{1}{2}\, \lambda_5 \left[ (\phi_1^\dagger \phi_2)^2 +
\textrm{h.c.} \right],\nonumber\\
V_S &=& M^2\, S^2 + \lambda_S\, S^4,\\
V_{S \phi}
&=&
z_1\,M (\phi_1^\dagger \phi_2)\, S + z_1^\ast\,M (\phi_2^\dagger \phi_1)\, S
+
\beta_1\, |\phi_1|^2\, S^2 + \beta_2\, |\phi_2|^2\, S^2.\nonumber
\end{eqnarray}
All parameters are real but $z_1$, so that CP violation in the scalar sector
is related with a complex $z_1$.

\begin{figure}[t]
\includegraphics*[height=3.5cm]{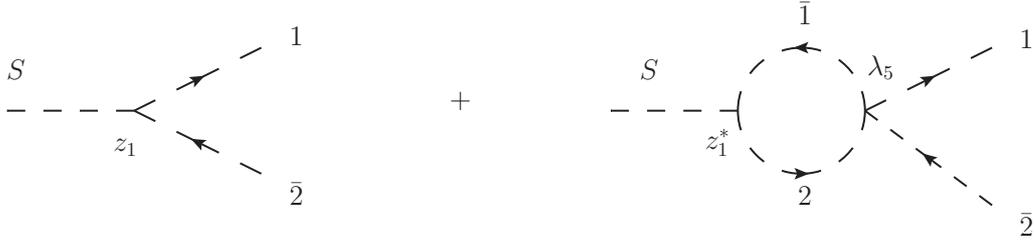}
\caption{\label{fig1}Diagrams contributing to the Higgs CP asymmetry. The
notation $1$ and $\bar{2}$ refers to $\phi_1$ and $\bar{\phi}_2$,
respectively.}
\end{figure}

We address now the question whether this model can be used to generate a CP
asymmetry in the Higgs sector that could be converted into a baryon
asymmetry. The basic idea is analogous to that of the standard leptogenesis
scenario. A population of $S$'s is produced through scattering processes at
temperatures $T \sim M \gg m_{\phi_1,\phi_2}$. This population decays away at
$T < M$, when the singlet scalar equilibrium density is Boltzmann suppressed.
If the interactions of the heavy singlet $S$ are CP-violating, and provided
that the relevant interactions are out of equilibrium, a net Higgs asymmetry
can be generated. The latter is then converted into a baryon asymmetry by the
sphalerons. To illustrate our mechanism, let us consider the tree-level and
one-loop diagrams\footnote{There is also a one-loop vertex correction to the
tree-level diagram, but since it carries the same phase $z_1$, it does not
contribute to the Higgs CP asymmetry.} depicted in Fig.~\ref{fig1}.

We have two contributions with different CP-odd phases; $z_1$ and $z_1^\ast$.
Because the second diagram is a loop diagram, a cut on this diagram leads to
an absorptive part that contains the CP-even phase needed for CP violation in
decays. As a result, the interference of the tree-level and one-loop
amplitudes leads to a nonvanishing CP asymmetry in the final Higgs states.
Defining this asymmetry as
\begin{equation} \label{CPasym}
\epsilon = \frac{\Gamma(S\rightarrow \phi_1\,\bar{\phi}_2)-\Gamma(S\rightarrow
\bar{\phi}_1\,\phi_2)}{ \Gamma(S\rightarrow
\phi_1\,\bar{\phi}_2)+\Gamma(S\rightarrow \bar{\phi}_1\,\phi_2)},
\end{equation}
we find
\begin{equation}
\epsilon \simeq\frac{\overline{\left|c_0 \mathcal{A}_0+c_1
\mathcal{A}_1\right|^2} -\overline{\left|c^\ast_0 \mathcal{A}_0+c^\ast_1
\mathcal{A}_1\right|^2}}{ \overline{\left|c_0 \mathcal{A}_0+c_1
\mathcal{A}_1\right|^2}+ \overline{\left|c^\ast_0 \mathcal{A}_0+c^\ast_1
\mathcal{A}_1\right|^2}} \simeq -4\frac{\text{Im}\left[c_0^\ast c_1\right]
\text{Im}\left[\overline{\mathcal{A}_0^\ast \mathcal{A}_1}\right]}{
2|c_0|^2\overline{|\mathcal{A}_0|^2}},
\end{equation}
where $\mathcal{A}_0$ and $\mathcal{A}_1$ are the tree-level and one-loop
amplitudes, respectively. For the decay of Fig.~\ref{fig1}, one has $c_0 = -
z_1$, $c_1= 3\,\lambda_5\, z_1^\ast$ and $\mathcal{A}_0=1$. Thus, the weak
phase gives
\begin{equation}
\text{Im}\left[c_0^\ast\, c_1\right]=- 3\, \text{Im}\left[z_1^{\ast}\,
\lambda_5\, z_1^{\ast} \right],
\end{equation}
while the strong phase comes from
\begin{equation}
\text{Im}\left[\overline{\mathcal{A}_0^\ast \mathcal{A}_1}\right]
=-\frac{1}{16\pi}.
\end{equation}
We then get
\begin{equation} \label{eps1}
\epsilon \simeq - \frac{3}{8\pi} \frac{\text{Im}\left[z_1^{\ast}\, \lambda_5\, z_1^{\ast} \right]}{2 |z_1|^2} =
\frac{3}{16\pi} \lambda_5 \sin[2\,\text{arg}(z_1)].
\end{equation}
Thus, in this simple scenario, the Higgs asymmetry is controlled by the
strength of the quartic parameter $\lambda_5$ and the phase of the coupling
$z_1$.

The final baryon asymmetry (baryon-to-entropy ratio) can be approximated as
\begin{equation}\label{YB}
  \mathcal{Y}_{\Delta B} \simeq \mathcal{Y}^{\rm eq}_S\times C\times\epsilon\,
  \eta,
\end{equation}
where the first factor is the equilibrium $S$ number density divided by the
entropy density, the second factor is the fraction of the Higgs asymmetry
converted into a baryon asymmetry by the sphalerons ($C=3/46$ in the present
case), and the efficiency factor $\eta\, (0 \leq \eta \leq 1)$ measures how
efficient the out-of-equilibrium $S$-decays are in producing the asymmetry.

Although a precise computation of $\eta$ requires the solution of a full set
of Boltzmann equations, simple analytical estimates can be given. It is useful
to introduce the decay parameter
\begin{equation}\label{Kappa}
  K=\frac{\Gamma_D(T=0)}{H(T=M)},
\end{equation}
where $\Gamma_D=|z_1|^2 M/(8\pi)$ is the tree-level decay rate of the singlet
$S$ into the two Higgs doublets. In the so-called weak washout regime ($K \ll
1$), i.e. when the scalar singlet decays strongly out of equilibrium, the
efficiency factor is $\eta\simeq 1$. In the strong washout regime ($K\gg 1$),
the efficiency does not depend on the initial conditions, and is mildly
suppressed as $\eta \simeq 1/K$. For intermediate values of $K$ ($K \lesssim
1$ or $K \gtrsim 1$) the efficiency depends on the assumed initial conditions.
We can roughly approximate it as $\eta \sim \min(1, 1/K)$, if $S$ has thermal
initial abundance, or $\eta \sim \min(K, 1/K)$, if $S$ has zero initial
abundance.

An estimate for $\mathcal{Y}_{\Delta B}$ can be obtained from
Eq.~\eqref{YB} in the form:
\begin{equation}\label{YBapp}
  \mathcal{Y}_{\Delta B} \simeq \frac{135 \zeta(3)}{92\pi^4 g_\ast}
  \epsilon\, \eta \simeq 2\times10^{-4}  \epsilon\, \eta .
\end{equation}
This is to be compared with the WMAP inferred value given in
Eq.~\eqref{YBobs}. Since Eq.~\eqref{eps1} leads to the upper bound $|\epsilon|
\lesssim 6\times 10^{-2} \lambda_5$, then Eq.~\eqref{YBapp}, when combined with
Eq.~\eqref{YBobs}, requires
\begin{equation}
\lambda_5 \, \eta \gtrsim 7\times 10^{-6}.
\end{equation}

Notice that between the mass scale $M$ and the EW scale an effective
quartic coupling $\lambda_5^\text{eff}=\lambda_5 +\tfrac{1}{2} z_1^2$ is
generated by $S$-exchange. Recalling that, for the relative asymmetry between
$\phi_1$ and $\phi_2$ to survive, interactions which exchange $\phi_1
\leftrightarrow \phi_2$ must be out of equilibrium until the electroweak
scale, then Eq.~\eqref{bdL} imposes $|\lambda_5^\text{eff}| \lesssim {\rm
few}\times 10^{-7}$. Thus, in this simple setup, unless there is a fine-tuned
cancellation between $\lambda_5$ and $z_1^2$ to satisfy this bound, we cannot
accommodate the observed baryon asymmetry~\eqref{YBobs}, even with a maximal
efficiency $\eta\simeq 1$.

\subsection{One extra singlet  and a third doublet}

We consider now a model with three doublet scalars and one real scalar singlet.
The Higgs doublets, $\phi_1$, $\phi_2$ and $\phi_3$, and the singlet scalar
$S$ transform under $Z_2$ as
\begin{equation}
\phi_1 \rightarrow - \phi_1,\quad
\phi_3 \rightarrow - \phi_3,\quad
\phi_2 \rightarrow \phi_2,\quad
S \rightarrow -S.
\label{Z2_3d1s}
\end{equation}
The singlet and the third doublet $\phi_3$  will be significantly heavier
than the EW scale; $\phi_1$ and $\phi_3$ are defined as the $Z_2$-odd mass
eigenstates. The $Z_2$-invariant Higgs potential can be written as
\begin{equation}
V_3 = V + \Delta V_\phi + \Delta V_{S \phi},
\label{V3}
\end{equation}
where $V$ is given in Eqs.~\eqref{VPPS}-\eqref{V_vp_Z2} and
\begin{eqnarray}\label{V3_vp_Z2}
\Delta V_\phi
&=&
m_{33}^2\, |\phi_3|^2
+ \, \tfrac{1}{2}\, \lambda_{3333}\, |\phi_3|^4
+ \lambda_{1133}\, |\phi_1|^2 |\phi_3|^2
+ \lambda_{2233}\, |\phi_2|^2 |\phi_3|^2
\nonumber\\
&&
 + \lambda_{1331}\, (\phi_1^\dagger \phi_3)(\phi_3^\dagger \phi_1)
+ \lambda_{2332}\, (\phi_2^\dagger \phi_3)(\phi_3^\dagger \phi_2)
\nonumber\\
&&
+  \left[\tfrac{1}{2}\lambda_{1313}(\phi_1^\dagger \phi_3)^2
+  \tfrac{1}{2}\lambda_{3232}(\phi_3^\dagger \phi_2)^2
+  \lambda_{1223}(\phi_1^\dagger \phi_2) (\phi_2^\dagger \phi_3)
+  \lambda_{1232}(\phi_1^\dagger \phi_2) (\phi_3^\dagger \phi_2) \right.\nonumber\\
&&
+  \left. \lambda_{1311}(\phi_1^\dagger \phi_3) (\phi_1^\dagger \phi_1)
+  \lambda_{1322}(\phi_1^\dagger \phi_3) (\phi_2^\dagger \phi_2)
+  \lambda_{1333}(\phi_1^\dagger \phi_3) (\phi_3^\dagger \phi_3) + \textrm{H.c.} \right]
 ,\nonumber\\
\Delta V_{S \phi}
&=&
z_3\,M (\phi_3^\dagger \phi_2)\, S + z_3^\ast\,M (\phi_2^\dagger \phi_3)\, S
+
\beta_3\, |\phi_3|^2\, S^2 + \left[\beta_{13}\, \phi_1^\dagger \phi_3 \, S^2 +
 \textrm{H.c.} \right].
\end{eqnarray}

The basic idea is similar to the 2HDM with an extra singlet. A population of
$S$'s is produced through scattering processes at temperatures $T \sim M >
m_{33} \gg m_{\phi_1,\phi_2}$. This population decays away at $T < M$, when
the singlet scalar equilibrium density is Boltzmann suppressed. If the
interactions of the heavy singlet $S$ are CP-violating, asymmetries among the
three Higgs doublets can be generated. When washout interactions are out of
equilibrium, the asymmetries can survive. The $\phi_3$ later decay to
$\phi_1$, leaving an asymmetry between $\phi_2$ and $\phi_1$. Assuming the
asymmetry from $S\to \phi_1 \phi_2^*$ to be negligible, due to the bounds on
$\lambda_5$ and $z_1^2$,  we neglect it and focus on a possible asymmetry
from $S\to \phi_3 \phi_2^*$.

We consider tree-level and one-loop diagrams analogous to those depicted in
Fig.~\ref{fig1}, with $\phi_1 \rightarrow \phi_3$, $z_1 \rightarrow z_3$, and
$\lambda_5 \rightarrow \lambda_{3232}$. The interference of the
tree-level and one-loop amplitudes leads to a nonvanishing CP asymmetry in
the final Higgs states:
\begin{equation} \label{CPasym3}
\epsilon = \frac{\Gamma(S\rightarrow \phi_3\,\bar{\phi}_2)-\Gamma(S\rightarrow
\bar{\phi}_3\,\phi_2)}{ \Gamma(S\rightarrow {\rm all})}
\simeq - \frac{3}{8\pi} \frac{\text{Im}\left[z_3^{\ast}\,
\lambda_{3232}\, z_3^{\ast} \right]}{2 |z_3|^2} = \frac{3}{16\pi} \lambda_{3232}
\sin[2\,\text{arg}(z_3)],
\end{equation}
where the contribution of $z_1$ to the total decay rate has been neglected.

For the relative asymmetry between $\phi_3 + \phi_1$ and $\phi_2$ to survive,
interactions which exchange $\phi_3 $ with $\phi_2$ must be out of
equilibrium until the $\phi_3$ decay. (Recall that we have already imposed
that interactions exchanging $\phi_1 \leftrightarrow \phi_2$ are out of
thermal equilibrium). By extrapolating Eq.~\eqref{bdL} to higher
temperatures, this imposes
\begin{equation}
|z_3|^2 , |\lambda_{3232}| <  {\rm few}\times
10^{-7} \sqrt{\frac {m_{33}}{\text{TeV}}}\,,
\end{equation}
assuming that $S$-exchange generates an effective
$\lambda_{3232}^\text{eff}=\lambda_{3232}+\tfrac{1}{2} z_3^2$ between the
temperature scales $M$ and $m_{33}$, and that $\phi_3$ decay at $T \simeq
10\, m_{33}$. From Eqs.~\eqref{YBapp} and \eqref{YBobs}, a large enough
asymmetry is obtained for $m_{33} \gtrsim 10^8$~GeV, provided the washout
effects are not too strong, $\eta \simeq \mathcal{O}(0.1)$. Since the
efficiency of Higgsogenesis is controlled by the decay parameter $K$ defined
in Eq.~\eqref{Kappa},
\begin{equation}
K= \frac{|z_3|^2}{136\pi}\frac{m_P}{M} \simeq 3 \times
\left(\frac{|z_3|^2}{10^{-4}}\right)\times\left(\frac{10^{12}\,\text{GeV}}{M}\right),
\end{equation}
this implies $M \gtrsim 10^{12}$~GeV.

\subsection{Two extra singlet scalars}

Let us now consider a model with two real scalar singlets $S_i\ (i=1,2)$, both
transforming under $Z_2$ as $S_i \rightarrow - S_i$. The Higgs potential can
be written as in Eqs.~\eqref{VPPS}-\eqref{V_vp_Z2}, but in this case $V_S$
and $V_{S \phi}$ contain additional terms. In particular, $V_{S \phi}$
contains the cubic terms $z_i\,M_i\, (\phi_1^\dagger \phi_2)\, S_i+
z_i^\ast\,M_i\, (\phi_2^\dagger \phi_1)\, S_i\,$, in addition to six quartic
terms. Similarly, $V_S$ has several quartic terms and, without loss of
generality, we choose a $\left\{ S_1, S_2\right\}$ basis where the quadratic
terms are already diagonalized, and assume $M_1 < M_2$.

\begin{figure}[t]
\includegraphics*[height=3.5cm]{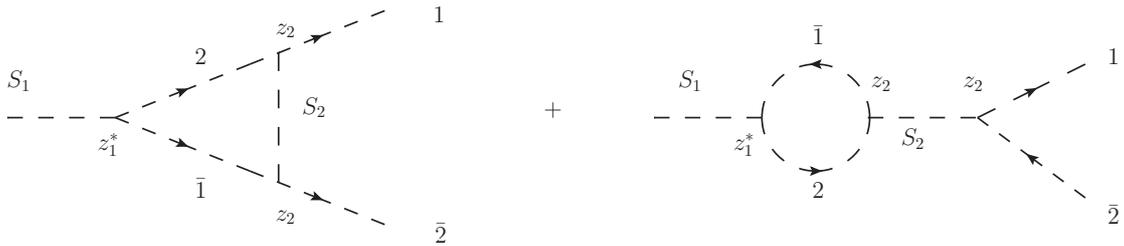}
\caption{\label{fig2}New diagrams contributing to the Higgs CP asymmetry in
the presence of two singlet scalars. The notation $1$ and $\bar{2}$ refers to
$\phi_1$ and $\bar{\phi}_2$, respectively.}
\end{figure}

In what follows, we make the following simplifying assumptions: The heavy
singlet spectrum is hierarchical, $M_1 \ll M_2$; there is a thermal production
of $S_1$ and negligible production of $S_2$. With these assumptions, the
Higgsogenesis mechanism will proceed via the out-of-equilibrium decays of
$S_1$. The decay $S_1 \rightarrow \phi_1 \bar{\phi}_2$ is still mediated by
the diagrams in Fig~\ref{fig1}, but we also have the additional (vertex and
self-energy) diagrams depicted in Fig.~\ref{fig2}.

These diagrams carry the phase of $z_1^\ast z_2^2$, which will be beaten
against the $z_1$ phase from the tree level diagram in Fig.~\ref{fig1}. As a
result, in this model, even if $\lambda_5$ vanishes, there is a CP-violating
contribution proportional to $\textrm{Im}(z_1^{\ast 2} z_2^2)$. The resulting
CP asymmetry, as given in Eq.~\eqref{CPasym}, can be evaluated following
the standard procedure. Neglecting the $\lambda_5$ contribution coming from
the one-loop diagram of Fig.~\ref{fig1}, we obtain
\begin{equation}
\epsilon_1 \simeq -\frac{1}{4\pi}
\frac{\text{Im}\left[(z^\ast_1z_2)^2\right]}{|z_1|^2}\left[\frac{1}{2}f(x_2)+g(x_2)\right],
\end{equation}
where $x_2 = M_2^2/M_1^2$, and
\begin{equation}
f(x)=x\ln \Big(\frac{x}{1+x}\Big),\quad g(x)=\frac{x}{1-x},
\end{equation}
are the vertex and self-energy one-loop functions, respectively.

In the hierarchical limit, $x_2\gg 1$, $f(x_2)\simeq -1$, $g(x_2) \simeq -1$,
and the CP asymmetry is approximately given by
\begin{equation} \label{eps2}
\epsilon_1 \simeq \frac{3}{8\pi}
\frac{\text{Im}\left[(z^\ast_1z_2)^2\right]}{|z_1|^2}=\frac{3}{8\pi} |z_2|^2
\sin[2\arg (z_1^\ast z_2)].
\end{equation}

From Eqs.~\eqref{YBapp}, \eqref{YBobs} and \eqref{eps2} we then conclude that
a successful generation of the baryon asymmetry within the present model
requires
\begin{equation}\label{z2constraint}
  |z_2|^2\, \eta \gtrsim {\rm few} \times 10^{-6},
\end{equation}
which in turn implies that $|z_2|^2 \gtrsim {\rm few} \times 10^{-6}$. Yet,
as in the case with one extra singlet, an effective quartic coupling
$\lambda_5^{\rm eff}=\lambda_5 + \tfrac{1}{2} z_1^2 +\tfrac{1}{2} z_2^2$ is
generated by $S$-exchange between the $M$ and EW temperature scales. The
latter should satisfy the bound $|\lambda_5^\text{eff}| \lesssim {\rm
few}\times 10^{-7}$. So, in this case, to accommodate the observed baryon
asymmetry~\eqref{YBobs} we should have some relation between $\lambda_5$,
$z_1^2$ and/or $z_2^2$; for example, $z_1 \simeq i z_2$ to avoid the
restrictive bounds on these couplings.

\section{Summary}
\label{sec:summary}

In this work, we have studied the possibility of generating the cosmological
baryon asymmetry in the context of 2HDM extensions of the SM, prior to the
electroweak phase transition. We have shown that if the Higgs-flavour
exchanging interactions are sufficiently slow in the early Universe, then a
relative asymmetry among the Higgs doublets corresponds to an effectively
conserved quantum number. Such a relative Higgs asymmetry can be transformed
into a baryon asymmetry by the sphalerons, without the need for $B-L$
violation.

Among the four possible types of $Z_2$ models considered, we have
demonstrated that this  ``split Higgsogenesis'' mechanism is only possible in
the framework of a type-I or type-X 2HDM. We then  presented simple scenarios
to generate a Higgs asymmetry,  based on inert type-I 2HDMs extended by heavy
singlet scalar fields and/or one extra Higgs doublet. In the presence of
CP-violating interactions, the out-of-equilibrium decays of the heavy
singlets into the Higgs doublets can produce a net Higgs asymmetry and the
mechanism of baryogenesis through (split) Higgsogenesis can be viable. Since
a successful implementation of our mechanism requires the scalar potential
parameters to satisfy definite bounds, we have also paid particular attention
to their basis-independent formulation.

\acknowledgments
 We are grateful to A. Barroso, M.~Krawczyk and R. Santos for
useful discussions. S.D. acknowledges partial support from the EU FP7 ITN
INVISIBLES (MC actions, PITN-GA-2011-289-442) and the Lyon Institute of
Origins. The work of R.G.F. and J.P.S. was partially supported by FCT -
\textit{Funda\c{c}\~{a}o para a Ci\^{e}ncia e a Tecnologia}, under the Projects
PEst-OE/FIS/UI0777/2011 and CERN/FP/123580/2011, and by the EU RTN Marie
Curie Project PITN-GA-2009-237920. The work of H.S. is funded by the European
FEDER and Spanish MINECO, under the Grant FPA2011-23596.

\appendix
\section{The mass basis and the top basis}
\label{sec:appendix}

Let us consider the quadratic terms of the scalar potential,
\begin{equation}
V_2 = \left( \phi_1^\dagger\ \ \phi_2^\dagger \right)\ \mathbf{M}
\left(
\begin{array}{c}
\phi_1\\
\phi_2
\end{array}
\right),
\end{equation}
where
\begin{equation}\label{matrix_M}
\mathbf{M}
=
\left(
\begin{array}{cc}
m_{11}^2 & -m_{12}^2\\*[2mm]
-\left(m_{12}^2\right)^\ast & m_{22}^2
\end{array}
\right)
\end{equation}
is a Hermitian matrix. Its eigenvalues are
\begin{equation}
M_{1,2}^2
=
\frac{m_{11}^2 + m_{22}^2}{2}
\pm
\frac{1}{2} \sqrt{(m_{11}^2 - m_{22}^2)^2 + 4 \left|m_{12}^2\right|^2}\,.
\label{M1sqM2sq}
\end{equation}

The top Yukawa couplings can be written as
\begin{equation}
- {\cal L}_Y
=
\bar{t}_L\
\left( y_1^t\ \ y_2^t \right)\,
\left(
\begin{array}{c}
\tilde{\phi}_1\\
\tilde{\phi}_2
\end{array}
\right)\ t_R
+ \textrm{H.c.},
\end{equation}
from which we may build the Hermitian matrix
\begin{equation}
\mathbf{H}
=
\left(
\begin{array}{cc}
|y_1^t|^2 & y_1^t\, y_2^{t \ast}\\*[2mm]
y_1^{t \ast}\, y_2^t & |y_2^t|^2
\end{array}
\right).
\end{equation}
Its eigenvalues are
\begin{equation}
\left(y^\textrm{top}\right)^2
=
|y_1^t|^2  + |y_2^t|^2,
\end{equation}
and zero.

In the mass basis, $\mathbf{M}$ is diagonal and $\mathbf{H}$ has the form
\begin{equation}
\mathbf{H}
=
\left(
\begin{array}{cc}
|y_1^{t {\rm m}}|^2 & y_1^{t{\rm m}}\, y_2^{t{\rm m} \ast}\\*[2mm]
y_1^{t{\rm m} \ast}\, y_2^{t{\rm m}} & |y_2^{t{\rm m}}|^2
\end{array}
\right).
\end{equation}
In the top basis, $\mathbf{H}$ is diagonal, only one Higgs scalar couples to
the top, and
\begin{equation}
\mathbf{M}
=
\left(
\begin{array}{cc}
m_{11}^{{\rm t}\, 2} & -m_{12}^{{\rm t}\, 2}\\*[2mm]
-\left(m_{12}^{{\rm t}\, 2}\right)^\ast & m_{22}^{{\rm t}\, 2}
\end{array}
\right).
\end{equation}
Notice that we have used the superscript ``m" to stress that the matrix
elements of $\mathbf{H}$ are to be calculated in the basis where $\mathbf{M}$
is diagonal. Similarly, the superscript ``t" indicates that the matrix
elements of $\mathbf{M}$ are to be calculated in the basis where $\mathbf{H}$
is diagonal.

There is physical content in the misalignment between the two bases, which
can be expressed in a basis-invariant way through the relation
\begin{equation}
\sin^2{\left( 2 \theta_{t} \right)}=
\frac{4\, \det\,[\mathbf{H}, \mathbf{M}]}{
\left[ 2\,\textrm{Tr}\,(\mathbf{H}^2) - (\textrm{Tr}\,\mathbf{H})^2 \right]
\left[2\,\textrm{Tr}\,(\mathbf{M}^2) - (\textrm{Tr}\,\mathbf{M})^2 \right]}\,.
\label{rel_angle}
\end{equation}
In the top basis, the above equation yields
\begin{equation}
\sin^2(2 \theta_{t})
=
\frac{4\, \left|m_{12}^{{\rm t}\,2}\right|^2}{
\left( m_{11}^{{\rm t}\, 2} - m_{22}^{{\rm t}\, 2} \right)^2
+ 4 \left|m_{12}^{{\rm t}\,2}\right|^2},
\label{theta_tM}
\end{equation}
or equivalently,
\begin{equation}
|\tan(2 \theta_{t})|=
\frac{2\,|m_{12}^{{\rm t}\,2}|}{|m_{11}^{{\rm t}\, 2} - m_{22}^{{\rm t}\, 2}|}.
\label{tantheta_tM}
\end{equation}
In the mass basis, it leads to
\begin{equation}
\sin^2(2 \theta_{t})
=
\frac{4\, \left|y_1^{t {\rm m}}\right|^2\, \left|y_2^{t {\rm m}}\right|^2}{
\left|y_1^{t {\rm m}}\right|^2 + \left|y_2^{t {\rm m}}\right|^2}\,.
\label{theta_Mt}
\end{equation}
For small angles we can then write
\begin{equation}
\theta_{t} \simeq \frac{\left|y_1^{t {\rm m}}\,y_2^{t {\rm m}}\right|}{
\sqrt{\left|y_1^{t {\rm m}}\right|^2 + \left|y_2^{t {\rm m}}\right|^2}} \simeq
\frac{\left|m_{12}^{{\rm t}\,2}\right|}{\left|m_{11}^{{\rm t}\, 2} -
m_{22}^{{\rm t}\, 2}\right|} \,. \label{thetasmall}
\end{equation}
This equation provides two different ways of writing the misalignment between
the top and the mass bases to be used in the text.

Thus far we did not need to specify whether the mass matrix $\mathbf{M}$ in
Eq.~\eqref{matrix_M} is to be calculated at a finite temperature $T$, or at
$T=0$; the expressions hold for any case. But one subtlety must be pointed
out when using Eq.~\eqref{thetasmall} with a temperature-dependent matrix
$\mathbf{M}(T)$. The thermal mass basis rotates as the temperature varies, as
can be seen from Eq.~(\ref{mtherm}). The top basis, on the other hand, is
temperature independent, and the bound of Eq.~(\ref{topbd}) must be satisfied
from the temperature when the Higgs asymmetry is created until the EWPT. This
implies that the thermal basis cannot rotate much during this period.


\begin{thebibliography}{99}
%
\bibitem{LHC}
 S.~Chatrchyan {\it et al.}  [CMS Collaboration],
  Phys.\ Lett.\ B {\bf 716} (2012) 30
  [arXiv:1207.7235 [hep-ex]];
 G.~Aad {\it et al.}  [ATLAS Collaboration],
  Phys.\ Lett.\ B {\bf 716} (2012) 1
  [arXiv:1207.7214 [hep-ex]].
%
\bibitem{Sakharov}
 A.~D.~Sakharov,
  Pisma Zh.\ Eksp.\ Teor.\ Fiz.\  {\bf 5} (1967) 32
   [JETP Lett.\  {\bf 5} (1967) 24]
   [Sov.\ Phys.\ Usp.\  {\bf 34} (1991) 392]
   [Usp.\ Fiz.\ Nauk {\bf 161} (1991) 61].
%
\bibitem{BAU}
 E.~W.~Kolb and S.~Wolfram,
  Nucl.\ Phys.\ B {\bf 172} (1980) 224
   [Erratum-ibid.\ B {\bf 195} (1982) 542];
  A.~D.~Dolgov,
  Phys.\ Rept.\  {\bf 222} (1992) 309.
\bibitem{BAU2} For recent reviews, see e.g.,
  S.~Davidson, E.~Nardi and Y.~Nir,
  Phys.\ Rept.\  {\bf 466} (2008) 105
  [arXiv:0802.2962 [hep-ph]];
  G.~C.~Branco, R.~Gonz\'{a}lez~Felipe and F.~R.~Joaquim,
  Rev.\ Mod.\ Phys.\  {\bf 84} (2012) 515
  [arXiv:1111.5332 [hep-ph]].
%

\bibitem{BAU@EPT}
V.~A.~Rubakov and M.~E.~Shaposhnikov,
  Usp.\ Fiz.\ Nauk {\bf 166} (1996) 493
   [Phys.\ Usp.\  {\bf 39} (1996) 461]
  [hep-ph/9603208];
 M.~Quiros,
  J.\ Phys.\ A {\bf 40} (2007) 6573;
 J.~M.~Cline,
  hep-ph/0609145.

\bibitem{hhg}
J.F.~Gunion, H.E.~Haber, G.~Kane and S.~Dawson,
{\it The Higgs Hunter's Guide} (Perseus Publishing,
Cambridge, MA, 1990).
%
\bibitem{2HDM}
For a recent review on 2HDM, see, for example,
G.C.~Branco, P.M.~Ferreira, L.~Lavoura, M.N.~Rebelo, M.~Sher and J.P.~Silva,
  Phys.\ Rept.\  \textbf{516}, 1 (2012)
  [arXiv:1106.0034 [hep-ph]].
%
\bibitem{sphalerons}
  P.~B.~Arnold and L.~D.~McLerran,
  Phys.\ Rev.\ D {\bf 36} (1987) 581;
  F.~R.~Klinkhamer and N.~S.~Manton,
  Phys.\ Rev.\ D {\bf 30} (1984) 2212.
%
\bibitem{Lavoura:1994fv}
   L.~Lavoura and J.~P.~Silva,
   Phys.\ Rev.\ D {\bf 50} (1994) 4619 [hep-ph/9404276];
   F.~J.~Botella and J.~P.~Silva,
Phys.\ Rev.\ D {\bf 51} (1995) 3870 [hep-ph/9411288].
%
\bibitem{DH}
  S.~Davidson and H.~E.~Haber,
  Phys.\ Rev.\ D {\bf 72} (2005) 035004
   [Erratum-ibid.\ D {\bf 72} (2005) 099902]
  [hep-ph/0504050].
%
\bibitem{KS}
  S.~Y.~.Khlebnikov and M.~E.~Shaposhnikov,
  Nucl.\ Phys.\ B {\bf 308} (1988) 885;
S.~Y.~.Khlebnikov and M.~E.~Shaposhnikov,
  Phys.\ Lett.\ B {\bf 387} (1996) 817
  [hep-ph/9607386];
J.~A.~Harvey and M.~S.~Turner,
  Phys.\ Rev.\ D {\bf 42} (1990) 3344.
%
\bibitem{reviews}
For recent reviews see, for example,
  B.~Grinstein and P.~Uttayarat,
arXiv:1304.0028 [hep-ph];  
  A.~Barroso, P.~M.~Ferreira, R.~Santos, M.~Sher and Jo\~{a}o P.~Silva,
arXiv:1304.5225 [hep-ph];
  C.~-Y.~Chen, S.~Dawson and M.~Sher,
arXiv:1305.1624 [hep-ph];
  O.~Eberhardt, U.~Nierste and M.~Wiebusch,
arXiv:1305.1649 [hep-ph];
  N.~Craig, J.~Galloway and S.~Thomas,
arXiv:1305.2424 [hep-ph];
P.~M.~Ferreira, R.~Santos, M.~Sher, and Jo\~{a}o P.~Silva,
arXiv:1305.4587 [hep-ph].
%
\bibitem{inert}
  N.~G.~Deshpande and E.~Ma,
Phys.\ Rev.\ D {\bf 18} (1978) 2574 ;
  R.~Barbieri, L.~J.~Hall and V.~S.~Rychkov,
  Phys.\ Rev.\ D {\bf 74} (2006) 015007 [hep-ph/0603188];
Q.~-H.~Cao, E.~Ma and G.~Rajasekaran,
Phys.\ Rev.\ D {\bf 76} (2007) 095011 [arXiv:0708.2939 [hep-ph]].
%
\bibitem{warsaw}
  M.~Krawczyk, D.~Sokolowska and B.~Swiezewska,
arXiv:1304.7757 [hep-ph];
  M.~Krawczyk, D.~Sokolowska, P.~Swaczyna and B.~Swiezewska,
arXiv:1305.6266 [hep-ph];
  B.~Swiezewska and M.~Krawczyk,
arXiv:1305.7356 [hep-ph];
  B.~Swiezewska,
arXiv:1209.5725 [hep-ph].
%
\bibitem{Aad:2013wqa}
  G.~Aad {\it et al.}  [ ATLAS Collaboration],
  arXiv:1307.1427 [hep-ex].
%
\bibitem{CMS_gg}
CMS Collaboration, document CMS-PAS-HIG-13-001 (2013).
%
\bibitem{CE}
  D.~Comelli and J.~R.~Espinosa,
  ``Bosonic thermal masses in supersymmetry,''
  Phys.\ Rev.\ D {\bf 55} (1997) 6253
  [hep-ph/9606438].
%
\bibitem{Ivanov}
  I.~P.~Ivanov,
  Acta Phys.\ Polon.\ B {\bf 40} (2009) 2789
  [arXiv:0812.4984 [hep-ph]];
 I.~F.~Ginzburg, I.~P.~Ivanov and K.~A.~Kanishev,
  Phys.\ Rev.\ D {\bf 81} (2010) 085031
  [arXiv:0911.2383 [hep-ph]].
%
\bibitem{ADMrev}
  H.~Davoudiasl and R.~N.~Mohapatra,
  New J.\ Phys.\  {\bf 14} (2012) 095011
  [arXiv:1203.1247 [hep-ph]];
%
  K.~Petraki and R.~R.~Volkas,
  arXiv:1305.4939 [hep-ph].
%
\bibitem{ST}
  G.~Servant and S.~Tulin,
  arXiv:1304.3464 [hep-ph].
%
\bibitem{BAUIDM}
  D.~Borah and J.~M.~Cline,
   Phys.\ Rev.\ D {\bf 86} (2012) 055001
   [arXiv:1204.4722 [hep-ph]];
%
  G.~Gil, P.~Chankowski and M.~Krawczyk,
   Phys.\ Lett.\ B {\bf 717} (2012) 396
   [arXiv:1207.0084 [hep-ph]];
%
  M.~Laine, G.~Nardini and K.~Rummukainen,
   JCAP {\bf 1301} (2013) 011
   [arXiv:1211.7344 [hep-ph]].
%
\bibitem{Komatsu:2010fb}
  E.~Komatsu {\it et al.}  [WMAP Collaboration],
  Astrophys.\ J.\ Suppl.\  {\bf 192} (2011) 18
  [arXiv:1001.4538 [astro-ph.CO]].
\end{thebibliography}
\end{document}